\documentclass[aps,pra,groupedaddress,superscriptaddress,twocolumn]{revtex4-1}

\usepackage[utf8x]{inputenc}
\usepackage{color}
\usepackage{bbm} 

\usepackage{amsfonts,amsmath,amssymb,stmaryrd}

\usepackage{graphicx}
\usepackage{subfigure}  
\usepackage{bbm}
\usepackage{hyperref}
\usepackage{epsfig}
\usepackage{mathrsfs}
\usepackage{verbatim}
\usepackage{ulem}	
\usepackage{siunitx}
\usepackage{soul} 

\usepackage[all]{hypcap}
\usepackage{braket}
\usepackage{array}

\DeclareSIUnit\BohrRadii{\mathrm{a}_0}

\newcommand{\ovr}{{\ensuremath{\overline{r}}}}
\newcommand{\ovq}{{\ensuremath{\overline{q}}}}
\newcommand{\Rb}{\ensuremath{^{87}\mathrm{Rb}}}
\newcommand{\cev}[1]{\reflectbox{\ensuremath{\vec{\reflectbox{\ensuremath{#1}}}}}}
\newcommand{\cg}[2]{\mathrm{C}^{#1}_{#2}}
\newcommand{\RMQN}{X}	
\newcommand{\FSQN}{X}	
\newcommand{\HFSQN}{Y}	

\begin{document}
\normalem	

\title{Photoassociation of rovibrational Rydberg molecules}

\author{O. Thomas}
\affiliation{Department of Physics and Research Center OPTIMAS, Technische Universität Kaiserslautern, 67663 Kaiserslautern, Germany}
\affiliation{Graduate School Materials Science in Mainz, 67663 Kaiserslautern, Germany}

\author{C. Lippe}
\affiliation{Department of Physics and Research Center OPTIMAS, Technische Universität Kaiserslautern, 67663 Kaiserslautern, Germany}

\author{T. Eichert}
\affiliation{Department of Physics and Research Center OPTIMAS, Technische Universität Kaiserslautern, 67663 Kaiserslautern, Germany}

\author{H. Ott}
\affiliation{Department of Physics and Research Center OPTIMAS, Technische Universität Kaiserslautern, 67663 Kaiserslautern, Germany}

\begin{abstract}

In this work we discuss the rotational structure of Rydberg molecules. We calculate the complete wave function in a laboratory fixed frame and derive the transition matrix elements for the photoassociation of free ground state atoms. We discuss the implications for the excitation of different rotational states as well as the shape of the angular nuclear wave function. We find a rather complex shape and unintuitive coupling strengths, depending on the angular momenta coupling that are relevant for the states. This work explains the different steps to calculate the wave functions and the transition matrix elements in a way, that they can be directly transferred to different molecular states, atomic species or molecular coupling cases.

\end{abstract}

\date{\today}

\maketitle

\section{Introduction}

Since their first theoretical description \cite{greene2000rydmols} and experimental discovery \cite{bendkowsky2009observation}, Rydberg molecules have been of great interest in theoretical as well as experimental studies. Rydberg molecules are formed by the low-energy scattering between an electron in a Rydberg state and a perturber atom in an electronic ground state. They have been investigated for a manifold of different excited states as well as for different atomic species \cite{bendkowsky2009observation, Booth2015CsMolecules, camargo2016strontiumRydMol}. Polyatomic states, with more than one perturber atom, have been found \cite{Bendowsky2010Trimer}. Also more exotic molecules, such as the trilobite and butterfly molecules with up to several kilodebye permanent electric dipole moment, have been produced in experiments \cite{Booth2015CsMolecules, niederprum2016observation}. Coherent control of the molecular states has been demonstrated \cite{butscher2010atom} and recently Rydberg molecules have been used as an Optical Feshbach resonance to tune the interactions in a many-body system \cite{sandor2017RydOFR, thomas2017experimentalOFR}.

The scattering interaction between the electron in a Rydberg state and the perturber atom is typically described by a Fermi type pseudo-potential taking into account S- and P-wave scattering. The calculation of the Born-Oppenheimer potentials in a molecular fixed frame is a well established task and has been described in numerous works \cite{hamilton2002rydmols, chibisov2002rydmols, Anderson2014angularmomentumcoupling, Eiles2017SpinEffectsRydMol, markson2016SpinEffectsRydMol}. However, a complete description of the molecular wave function in a laboratory fixed frame and the calculation of transition matrix elements for photoassociation to different rotational states was missing so far. This task will be undertaken in this work. While we will concentrate on the class of ultra-long range Rydberg molecules most arguments are also valid for butterfly and trilobite molecules.


The outline of this article is as follows: In section \ref{sec:RydMols}, the calculation of the Born-Oppenheimer potentials and the description of diatomic Rydberg molecules is briefly revised. We will discuss the different angular momentum couplings and introduce the full molecular wave function in a laboratory fixed frame. In section \ref{sec:GroundState}, the wave function of two free ground state atoms shall be introduced and transformed into the same basis as the Rydberg molecule. In section \ref{sec:TransitionElements} we will then calculate the transition matrix elements for one and two photon transitions and, finally, we discuss the implications for the excited rotational states in section \ref{sec:RotationalStates}.

\begin{figure}
	\includegraphics[width=1\columnwidth]{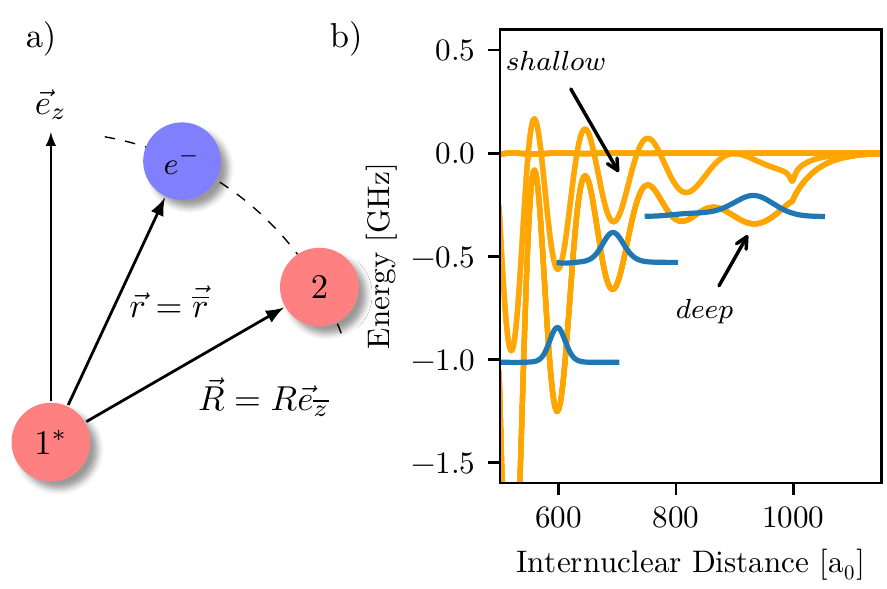}
	\caption{\label{fig:RydMolPotential} a) Coordinate system used throughout this work. Atom $1$ is excited to a Rydberg state and atom $2$ is bound to a molecular state by the scattering interaction with the electron in the Rydberg state. b) Born-Oppenheimer potential curves $U_\mathrm{BO}(R)$ for \Rb{} around the $(25\mathrm{P}_{3/2}+5\mathrm{S}_{1/2},F=2)$ manifold. Each potential type is degenerate with different $\Omega$ quantum numbers. The vibrational ground state in each well and the corresponding radial nuclear wave function $\mathscr{F}(R)$ for the so-called $deep$ potential is depicted in blue.}
\end{figure}

Before starting let us first introduce the coordinate system used throughout this work (see figure \ref{fig:RydMolPotential}a). We will consider two atoms with one of them being excited into a Rydberg state. The coordinate of the excited electron with respect to its positively charged nucleus will be labelled with $\vec r$ and the separation between the two nuclei with $\vec R$. In a molecular fixed frame we will choose the $z$-axis to coincide with the symmetry axis of the system, thus only having to consider the absolute value $R$. To distinguish the electronic coordinate in the two systems we will overline it in the molecular frame. Thus $\vec\ovr$ is the electronic coordinate in the molecular frame and $\vec r$ in the laboratory fixed frame. To transform the coordinate systems into each other we will use the three Euler angles $\alpha$, $\beta$ and $\gamma$ in z-y-z convention. The first two Euler angles may be identified as the usual coordinates $\Phi(=\alpha)$ and $\Theta(=\beta)$ in spherical coordinates, while the third coordinate corresponds to a rotation around the symmetry axis. While this may be ambiguous for a diatomic molecule we will nevertheless keep it for the sake of generality. As we will often use Wigner-D matrices in this work to transform from one frame to the other and to describe rotational states it should be noted that we will stick to the convention used by Rose et al. \cite{rose1957elementary}.

\section{Rydberg molecules}
\label{sec:RydMols}

Diatomic Rydberg molecules are typically described in a molecular fixed frame by a set of quantum numbers including all involved spins except the nuclear spin of the Rydberg atom, which is neglected due to the small hyperfine interaction in Rydberg atoms. In an atomic basis, the quantum numbers are thus $\ket{n_1 l_1 j_1 \overline{m}_{j_1}}\otimes\ket{n_2 l_2 j_2 \overline{m}_{j_2} I_2 \overline{m}_{I_2}}$ with the index $1$ indicating the atom excited to the Rydberg state and $2$ the bound perturber atom. We have dropped $s_1=s_2=1/2$ for simplicity. The bars indicate, that the atomic basis states are taken with respect to the internuclear axis. The electronic Hamiltonian in Born-Oppenheimer approximation is then written as \cite{Anderson2014angularmomentumcoupling}:
\begin{align}
	\hat H(\vec\ovr, \vec R) = &-\frac{\hbar^2}{2 m_e}\nabla^2_\ovr +V_C(\ovr) + V_\mathrm{fs}\nonumber\\ 
	&+ \sum_{i=S,T} 2\pi A_s^i(k)\delta^3(\vec\ovr - \vec R) \hat{\mathbbm{P}}_i\nonumber\\
	&+ \sum_{i=S,T} 6\pi A_p^i(k)\delta^3(\vec\ovr - \vec R) \cev\nabla\cdot\vec\nabla \,\hat{\mathbbm{P}}_i\nonumber\\
	&+A_\mathrm{hfs} \hat S_2\cdot\hat I_2.
	\label{equ:hamiltonian}
\end{align}
Here, the first three terms correspond to the kinetic energy of the Rydberg electron, its Coulomb interaction with the ionic core and its fine structure. The next two terms are the S- and P-wave interaction for singlet and triplet scattering of the Rydberg electron and the perturber atom, with $A_{s/p}^i(k)$ the energy dependent S-/P-wave scattering length and $\hat{\mathbbm{P}}_{i}$ the singlet/triplet projector for $i=S/T$, respectively. The last term corresponds to the hyperfine interaction in the perturber atom.

\begin{figure}
	\includegraphics[width=1\columnwidth]{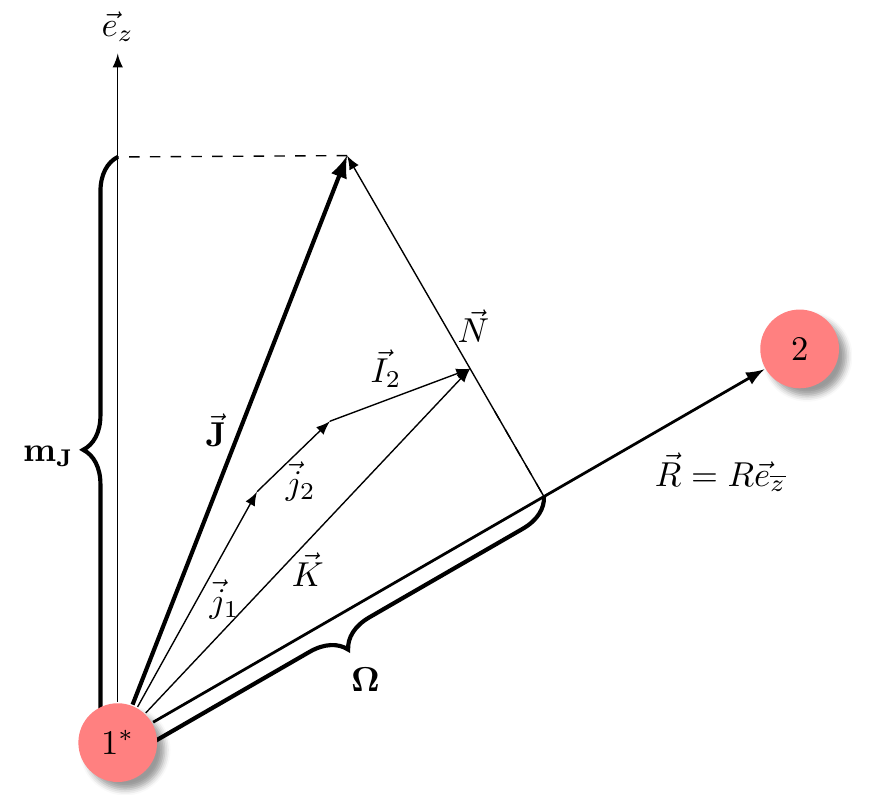}
	\caption{\label{fig:AngularCoupling} Hund's case c) description for Rydberg molecules. If the rotational energy is small compared to the scattering interaction, which itself is small compared to the fine and hyperfine structure splitting, a Hund's case c) is the best description for a Rydberg molecule. Within this the angular momenta $\vec j_1$, $\vec j_2$ and $\vec I_2$ are first coupled to the angular momentum $\vec K$. Its projection onto the internuclear axis is denoted by $\Omega$, which couples with the nuclear orbital angular momentum $\vec N$ of the two nuclei to the total angular momentum $\vec J = \vec N + \Omega\vec e_{\overline{z}}$. The good quantum numbers for the system are then $J$, it's projection on to the laboratory fixed axis $m_J$, and $\Omega$.}
\end{figure}

The solution of the corresponding Schrödinger equation to eq.\,\eqref{equ:hamiltonian} leads to Born-Oppenheimer potential curves $U_\mathrm{BO}(R)$ and electronic eigenstates $\ket{\Phi_\mathrm{mol}(R)}$, which parametrically depend on the internuclear distance. The calculation of the potentials and eigenstates has been described in detail elsewhere \cite{hamilton2002rydmols, chibisov2002rydmols, Anderson2014angularmomentumcoupling, Eiles2017SpinEffectsRydMol, markson2016SpinEffectsRydMol}. Here, we will assume that $U_\mathrm{BO}(R)$ and 
\begin{equation}
\begin{split}
	\ket{\Phi_\mathrm{mol}(R)}& = \sum_i c_i(R) \\
	&\times \left[\ket{n_1 l_1 j_1 \overline{m}_{j_1}}\otimes\ket{n_2 l_2 j_2 \overline{m}_{j_2} I_2 \overline{m}_{I_2}}\right]_i
\end{split}
\end{equation} are known in the molecular fixed frame. The coefficients $c_i$ depend on $R$ and express the electronic configuration in the atomic basis, fulfilling the normalization condition $\sum_i c_i(R)^2 = 1$. Note that the index $i$ runs over all atomic basis states.

In figure \ref{fig:RydMolPotential}b) the potential curves around the $(25\mathrm{P}_{3/2} + 5\mathrm{S}_{1/2}, F=2)$ manifold in \Rb{} are depicted. The composition of the different potential curves is rather complex and depends on the relative strength of the fine structure, hyperfine structure and scattering interaction. In the present example, three potential curves occur. We will discuss in this article states in the so-called $deep$ and $shallow$ potential, as in most cases both of them support bound molecular states and are the appropriate description in all species having a hyperfine structure in the perturber atom exceeding the scattering interaction.

The rotational energy of Rydberg molecules is typically smaller than any of the other energy scales, as the rotational constant $B=\hbar^2/2\mu d^2$ is very small for states with large bond length $d$ ($B=h\times\SI{46}{kHz}$ for \Rb{} with $d=950\si{a_0}$). Thus, the angular momentum coupling for most Rydberg molecules can be best described either in a Hund's case a) or c) depending on the size of the fine structure splitting, hyperfine structure splitting and scattering interaction \cite{Anderson2014angularmomentumcoupling}. An exception are ultra-long range Rydberg molecules with S-states, as they possess no fine structure and are thus best described with Hund's case b). We will discuss in section \ref{sec:RotationalStates} how to handle this.

In this work, we will base our description on a Hund's case c) description, for which the fine structure splitting and the hyperfine structure splitting dominate the scattering interaction. This is also the most intuitive choice for the chosen basis to solve eq.\,\eqref{equ:hamiltonian}. Note that because the basis states for each Hund's case are complete, the overall result doesn't depend on the chosen basis and we could base our description on any one.

In Hund's case c), the two electronic angular momenta and the nuclear spin of atom $2$ are coupled to the angular momentum $\vec K = \vec j_1 + \vec j_2 + \vec I_2$, with projection $\Omega$ onto the internuclear axis (see figure \ref{fig:AngularCoupling}). This projection is then coupled with the nuclear orbital angular momentum $\vec N$ to form the total angular momentum $\vec J = \Omega \vec e_{\overline{z}} + \vec N$, with projection $m_J$ onto the laboratory fixed $z$-axis. The rotational energy of such a state is given by \cite{bransden2003physics}
\begin{equation}
	E_\mathrm{rot} = B[J(J+1)-\Omega^2].
	\label{equ:RotEnergy}
\end{equation}
It should be noted that $K$ is not a good quantum number in a case a) description, however its projection onto the internuclear axis $\Omega$ will always be a good quantum number, because of the symmetry of the system.

To completely characterize a molecular state we use $\ket{\RMQN, \nu, J, m_J, \Omega}$, where we have summarized all electronic quantum numbers (except $\Omega$) into one single label $\RMQN$. The different vibrational states, characterized by $\nu$, are obtained solving the nuclear Schrödinger equation in Born-Oppenheimer approximation:
\begin{equation}
	\left( -\frac{\hbar^2}{2\mu}\frac{d^2}{dR^2} + U_\mathrm{BO}^{\RMQN,\Omega}(R) \right) \mathscr{F}^{\RMQN,\Omega}_\nu(R) = E_\mathrm{mol} \mathscr{F}^{\RMQN,\Omega}_\nu(R)
\end{equation}
where $\mu$ is the reduced mass of the system. We will assume that this equation is solved either numerically or in some analytic way to obtain the molecular energy $E_\mathrm{mol}$ and the radial nuclear wave function $\mathscr{F}^{\RMQN,\Omega}_\nu(R)$. Note that $E_\mathrm{mol}$ doesn't include the rotational energy eq.\,\eqref{equ:RotEnergy} yet.

Altogether, the state vector of a diatomic Rydberg molecule can be written as \cite{bransden2003physics, rose1957elementary}:
\begin{equation}
	\ket{\Psi_\mathrm{mol}(\vec R)} = \frac{1}{R}\mathscr{F}^{\RMQN,\Omega}_\nu(R) \, \mathscr{H}^J_{m_J,\Omega}(\alpha,\beta,\gamma) \, \ket{\RMQN,\Omega}(R)
	\label{equ:RydMolWF}
\end{equation}
where $\mathscr{H}^J_{m_J,\Omega} = \sqrt{\frac{2J+1}{8\pi^2}}D^{J*}_{m_J,\Omega}$ are normalized Wigner-D matrices describing the orientation of the molecular axis in the laboratory frame. Here, we have expressed the electronic state $\ket{\Phi_\mathrm{mol}(R)}$ through the molecular basis $\ket{\RMQN,\Omega}(R)$. Note that the possible values of the total angular momentum $J$ is restricted by the conditions $J\geq\Omega$ and the usual $J\geq m_J$.
We will use this expression later on to calculate transition matrix elements for the photoassociation of two ground state atoms. 


\section{Initial state}
\label{sec:GroundState}

In ultracold gases, the initial state for photoassociation usually corresponds to two free ground state atoms undergoing an S-wave collision. The formulas discussed above are not easily adapted for this case as an internuclear axis cannot be defined for spherical symmetry. In a molecular language, such a situation is best described in the often neglected Hund's case e). Therefore, in order to calculate transition matrix elements, we have to transform the wave function of the two free ground state atoms into a molecular frame.

To this purpose, we assume that both atoms are initially prepared in a hyperfine ground state $\ket{n_i l_i j_i f_i m_{f_i}}$ ($s_i=1/2$ as before) where $i=1,2$ labels the two atoms. In a typical experiment this initial state will be prepared with respect to the laboratory frame and we thus do not overline the magnetic quantum numbers. As before, we separate the nuclear spin of one atom
\begin{equation}
	\ket{\HFSQN_1 f_1 m_{f_1}} = \sum_{m_{j_1}, m_{I_1}} \cg{j_1,I_1,f_1}{m_{j_1},m_{I_1},m_{f_1}} \ket{\FSQN_1 j_1 m_{j_1}}\ket{I_1 m_{I_1}}
\end{equation}
where $\cg{j_1,I_1,f_1}{m_{j_1},m_{I_1},m_{f_1}}$ is a Clebsch-Gordon coefficient and we have summarized the remaining quantum numbers in $\HFSQN_1$ and $\FSQN_1$ for the hyperfine and fine structure case, respectively. We then couple the fine structure state of atom 1 with the hyperfine state of atom 2 to the total angular momentum $\vec K = \vec j_1 + \vec f_2$:
\begin{equation}
\begin{split}
	&\ket{\HFSQN_1 f_1 m_{f_1}}\ket{\HFSQN_2 f_2 m_{f_2}} \\
	&= \sum_{m_{j_1}, m_{I_1}} \cg{j_1,I_1,f_1}{m_{j_1},m_{I_1},m_{f_1}} \ket{\FSQN_1 j_1 m_{j_1}}\ket{I_1 m_{I_1}}\ket{\HFSQN_2 f_2 m_{f_2}}\\
	&= \sum_{m_{j_1}, m_{I_1}, K} \cg{j_1,I_1,f_1}{m_{j_1},m_{I_1},m_{f_1}}\cg{j_1,f_2,K}{m_{j_1},m_{f_2},m_{K}} \\&\quad\times\ket{\FSQN_1 \HFSQN_2 j_1 f_2 K m_{K}}  \ket{I_1 m_{I_1}}
\end{split}
\end{equation}
with $m_K = m_{j_1}+m_{f_2}$ the projection quantum number along the laboratory fixed $z$-axis. We transform this coupled state into a molecular fixed frame using \cite{rose1957elementary}:
\begin{equation}
	\ket{\FSQN_1 \HFSQN_2 j_1 f_2 K m_K} = \sum_{\Omega} D^{K*}_{m_K,\Omega}(\alpha, \beta, \gamma)\ket{\FSQN_1 \HFSQN_2 j_1 f_2 K \Omega}.
\end{equation}
So far we have not yet included the rotational state of the two nuclei. While for most ultracold experiments usually only S-wave scattering has to be considered, for Rydberg molecules higher partial waves might be of relevance because of the long bond length in Rydberg molecules. As a reference, for $r=\SI{2000}{\BohrRadii}$ the P-wave centrifugal barrier for \Rb{} is only \SI{1}{\micro\K} high. Therefore, in the temperature range of a typical magneto-optical trap higher partial waves cannot be neglected.

We represent the angular part of the scattering state by the relative angular momentum $N$ of the two nuclei:
\begin{equation}
	\braket{\vec R|N,m_N} = \sqrt{\frac{2N+1}{8\pi^2}} D^{N*}_{m_N,0}(\alpha,\beta,\gamma),
\end{equation}
which is equivalent to the more commonly used spherical harmonics, as $N$ is an integer. Using the properties of the Wigner-D matrices we can combine the two separate parts and end up with the total angular momentum $\vec J = \vec K + \vec N$ and the angular momentum projection $m_J = m_K+m_N$. This leads to the total angular momentum wave function expressed in the molecular frame:
\begin{widetext}
\begin{equation}
\begin{split}
	&\ket{\HFSQN_1 f_1 m_{f_1}}\ket{\HFSQN_2 f_2 m_{f_2}}\braket{\vec R|N m_N}\\
	&=\sum_{m_{j_1}, m_{I_1}, K, \Omega, J} \sqrt{\frac{2N+1}{8\pi^2}} \cg{j_1,I_1,f_1}{m_{j_1},m_{I_1},m_{f_1}}\cg{j_1,f_2,K}{m_{j_1},m_{f_2},m_{K}}
\cg{K,N,J}{m_{K},m_{N},m_{J}}\cg{K,N,J}{\Omega,0,\Omega} D^{J*}_{m_J,\Omega}(\alpha,\beta,\gamma)\ket{\FSQN_1 \HFSQN_2 j_1 f_2 K \Omega}\ket{I_1 m_{I_1}}.
\end{split}
\label{equ:GroundWF}
\end{equation}
\end{widetext}
An atomic pair scattering in a single channel characterized by $\ket{N, m_N}$ thus corresponds to a sum over different total angular momenta $\vec J$ and different angular projections $\Omega$. It should also be noted, that $m_J$ has no well defined value in above description. Only $m_J + m_{I_1}$ as the total angular momentum projection number has a fixed value.

To obtain a full wave function as written in equation \eqref{equ:RydMolWF} for the excited state we are only missing the relative radial wave function $\mathscr{F}_g(R)$. This might be some free scattering state or a relative wave function of two atoms bound in a potential depending on the system investigated. We will assume that $\mathscr{F}_g(R)$ is some experimentally given function and can then write the total wave function as before. To simplify from here on we will assume that the two atoms are in fully stretched states colliding in an s-wave ($N=0$). For this case all Clebsch-Gordon coefficients in eq.\,\eqref{equ:GroundWF} are equal to one and we can simply write the total wave function as:
\begin{equation}
\begin{split}
	\ket{\Psi_{g}(\vec R)} = \frac{1}{R}\mathscr{F}_g(R) &\sum_{\Omega=-J}^{J} \sqrt{\frac{1}{8\pi^2}}D^{J*}_{m_J,\Omega}(\alpha,\beta,\gamma)\\
		&\times\ket{\FSQN_1 \HFSQN_2 j_1 f_2 K \Omega}\otimes\ket{I_1 m_{I_1}}.
\end{split}
\label{equ:GroundWFstretched}
\end{equation}
This will be our starting point to calculate transition matrix elements for photoassociation of molecular Rydberg states.

\section{Transition matrix elements}
\label{sec:TransitionElements}

To calculate the transition matrix elements to different molecular states we first consider a single photon transition, where the polarization of the photon is given in the laboratory fixed frame. The dipole operators, which have to be evaluated, are therefore $d_q$ with $q=0,\pm1$ for $\pi, \sigma^\pm$ transitions, respectively. As we have expressed the electronic wave functions in a molecular fixed frame, we also have to transform the dipole operator accordingly \cite{rose1957elementary}
\begin{equation}
	\hat d_q = \sum_\ovq D^{1*}_{q,\ovq} \hat d_\ovq
\end{equation}
where $\hat d_\ovq$ is the dipole operator in the molecular fixed frame. In the following, we assume that only atom 1 can be excited and evaluate $\hat{d_q}_1\otimes\mathbbm{1}_2$. If both atoms are in the same initial state and can be excited, one has to evaluate $\hat{d_q}_1\otimes\mathbbm{1}_2 + \mathbbm{1}_1\otimes\hat{d_q}_2$, in order to account for the correct symmetry of the system. An excitation will then couple to a superposition of atom $1$ excited and atom $2$ bound in its Rydberg electron wave function and vice versa. In total this leads to an additional factor of $\sqrt{2}$ which can be taken into account by hand if necessary.

Priming all quantum numbers associated with the molecular state and dropping the nuclear wave function of atom $1$, the transition matrix element  is given as:
\begin{equation}
\begin{split}
	&\bra{\Psi_\mathrm{mol}(\vec R)} \hat d_{q1}\otimes\mathbbm{1}_2 \ket{\Psi_{g}(\vec R)} = \\
	&\frac{\sqrt{2J'+1}}{8\pi^2} \int dR \int d\alpha \sin\beta d\beta d\gamma \\
	&\times\mathscr{F}^{\RMQN',\Omega'*}_{\nu'}(R)\mathscr{F}_g(R) \\
	&\times\sum_{\overline{q}, \Omega} \Big[ D^{J'}_{m_J',\Omega'}(\alpha,\beta,\gamma) D^{1*}_{q,\overline{q}}(\alpha,\beta,\gamma) D^{J*}_{m_J,\Omega}(\alpha,\beta,\gamma)\\
	&\times\bra{\RMQN',\Omega'} d_{\overline{q}1}\otimes\mathbbm{1}_2 \ket{\FSQN_1 \HFSQN_2 j_1 f_2 K \Omega}(R)\Big].
\end{split}
\label{equ:MoleculeDipoleMatrixElement}
\end{equation}
This equation is the central result of this work. For ultra-long range molecules the parametric dependency on $R$ of the excited electronic state is very weak. We can thus use the Frank-Condon principle and split equation \eqref{equ:MoleculeDipoleMatrixElement} into three separate parts. In this way, we only have to take care of the summations over $\Omega$ and $\overline{q}$ later on.

The first term is a Frank-Condon integral for the radial wave function 
\begin{equation}
	\int dR \mathscr{F}^{\RMQN',\Omega'*}_{\nu'}(R)\mathscr{F}_g(R).
	\label{equ:FrankCondonFactor}
\end{equation}
As we have assumed that the molecular wave function and the ground state wave function are known, the integral can be calculated straightforward. It should be noted that because of the large bond length of Rydberg molecules this factor is typically much larger than for conventional molecules.

The second term can be understood as a H\"onl-London factor and can be calculated analytically
\begin{equation}
\begin{split}
	&\frac{\sqrt{2J'+1}}{8\pi^2} \int d\alpha \sin\beta d\beta d\gamma D^{J'}_{m_J',\Omega'} D^{1*}_{q,\ovq} D^{J*}_{m_J,\Omega}\\
	&= \frac{(-1)^{m_J'-\Omega'+q-\ovq+m_J-\Omega}}{\sqrt{2J'+1}} \cg{J,1,J'}{-m_J,-q,-m_J'} \cg{J,1,J'}{-\Omega,-\ovq,-\Omega'}.
\end{split}
\end{equation}
From this expression, the selection rules for the dipole transition can be extracted. The projection of the total angular momentum $J$ onto the laboratory fixed axis $m_J$, is changed by $q$, $\Delta m = m_J' - m_J = q$ and the total angular momentum $J$ can change at most by one quantum, $\Delta J = J'-J=0,\pm 1$. Regarding the quantum number $\Omega$, the selection rule is $\Delta \Omega = \Omega' - \Omega = \ovq$. However, the situation is more complex because the initial state is a superposition of all $\Omega$, with $|\Omega|\leq J$, and in the molecular frame, all transitions $\overline{q}$ with respect to the internuclear axis are in principle allowed. As a consequence, all quantum numbers $\Omega'$ with $|\Omega'| \leq J+1$ can be excited.

The last term corresponds to the electronic transition matrix element:
\begin{equation}
\begin{split}
	\bra{\RMQN',\Omega'} d_{\overline{q}1}\otimes\mathbbm{1}_2 \ket{\FSQN_1 \HFSQN_2 j_1 f_2 K \Omega}\\
	= \sum_i c_i \left[\bra{n_1' l_1' j_1' \overline{m}_{j_1}'}\otimes\bra{n_2' l_2' j_2' \overline{m}_{j_2}' I_2' \overline{m}_{I_2}'}\right]_i \\
	\times (d_{\overline{q}1}\otimes\mathbbm{1}_2) \ket{\FSQN_1 \HFSQN_2 j_1 f_2 K \Omega}.
\end{split}
\end{equation}
To calculate it explicitly we have to decouple the angular momentum of the initial state given in the molecular frame and write it in the atomic basis with internuclear axis as quantization axis
\begin{equation}
\begin{split}
	&\ket{\FSQN_1 \HFSQN_2 j_1 f_2 K \Omega} = \sum_{\overline{m}_{j_1},\overline{m}_{f_2},\overline{m}_{j_2},\overline{m}_{I_2}} \cg{j_1,f_2,K}{\overline{m}_{j_1},\overline{m}_{f_2},\Omega}\cg{j_2,I_2,f_2}{\overline{m}_{j_2},\overline{m}_{I_2},\overline{m}_{f_2}}\\
	&\quad\times\ket{n_1,l_1,j_1,\overline{m}_{j_1}}\otimes\ket{n_2,l_2,j_2,\overline{m}_{j_2},I_2,\overline{m}_{I_2}}.
\end{split}
\label{equ:GroundStateDecoupling}
\end{equation}
This transition matrix element is now completely written in the molecular frame.
From here on, the dipole matrix elements can be calculated straightforward, as has been done in various other works \cite{theodosiou1984lifetimeRyd, gallagher2005rydberg}. They are subject to the usual selection rules of an atomic transition $\Delta j_1=0,\pm1, \Delta \overline{m}_{j1} = 0,\pm1$, while maintaining the quantum numbers in atom $2$, $\Delta n_2 = \Delta l_2 = \Delta \overline{m}_{j_2} = \Delta I_2 = \Delta \overline{m}_{I_2} = 0$.

With this formalism at hand, eq.\,\eqref{equ:MoleculeDipoleMatrixElement} can be evaluated to get the transition matrix elements and the corresponding Rabi frequencies to excite a particular molecular Rydberg state.

Before discussing the implications on the excited nuclear wave functions and their orientation in space, we want to briefly discuss how to handle the experimentally more common two photon transitions and the implications of the neglected hyperfine interaction in atom $1$. 

For a two photon transition the scheme is not more complicated than discussed so far. We can simply replace the initial state with the intermediate state $\ket{ig}$, where $i$ denotes the intermediate level of atom 1 and evaluate equation \eqref{equ:MoleculeDipoleMatrixElement} to get the transition matrix element $d_\mathrm{mol}$. Because for typical experiments the detuning from the intermediate state $\Delta$ is larger than its hyperfine splitting in most cases a description of $\ket{i}$ in a fine structure picture will be sufficient. 

The total coupling strength is then $\propto d_\mathrm{lower}d_\mathrm{mol}/(2\Delta)$ where $d_\mathrm{lower}$ is the transition matrix element for the lower transition.

Neglecting the hyperfine interaction in the Rydberg atom is a good approximation in the case of Rydberg states with $l>0$ as for these the hyperfine interaction is smaller then the rotational energy. For S-states however the hyperfine interaction will exceed in most cases the rotational energy ($\nu_\mathrm{hfs} \approx \SI{33}{GHz} (n^*)^{-3}$ in \Rb{} \cite{Li2003MMspectroscopy}). The correct coupling  would therefore require to include the hyperfine interaction before adding the nuclear rotation. This will lead to some changes in the above description, which should be straightforward to include.


\section{Rotational states of Rydberg molecules}
\label{sec:RotationalStates}

To discuss the angular nuclear wave function of Rydberg molecules, we have to give up the generality we have tried to keep so far. In the following, we will give a few selected examples, which capture the most relevant aspects of the photoassociation process. We will discuss molecular states in \Rb{} bound in the outermost well of the molecular $deep$ and $shallow$ potentials. The eigenvectors $\ket{\RMQN' \Omega'}$ are retrieved from a full numerical diagonalization of equation \eqref{equ:hamiltonian} at the minimum of the potential well. To further simplify the treatment, we will assume that the two atoms are colliding in an S-wave. To check the numerical procedure for consistency, we have verified that in the limit of vanishing molecular interaction, the formalism leads to a spherical nuclear wave function in the excited state. It should also be noted that despite using some fixed main quantum numbers $n$ and $n'$ for the ground, intermediate and excited state, the principle arguments will be valid for any main quantum number. Calculated transition matrix elements from eq.\,\eqref{equ:MoleculeDipoleMatrixElement} will be given in reference to the equivalent atomic transition, however, excluding the Frank-Condon factor eq.\,\eqref{equ:FrankCondonFactor}.

\begin{figure}
	\includegraphics[width=1\columnwidth]{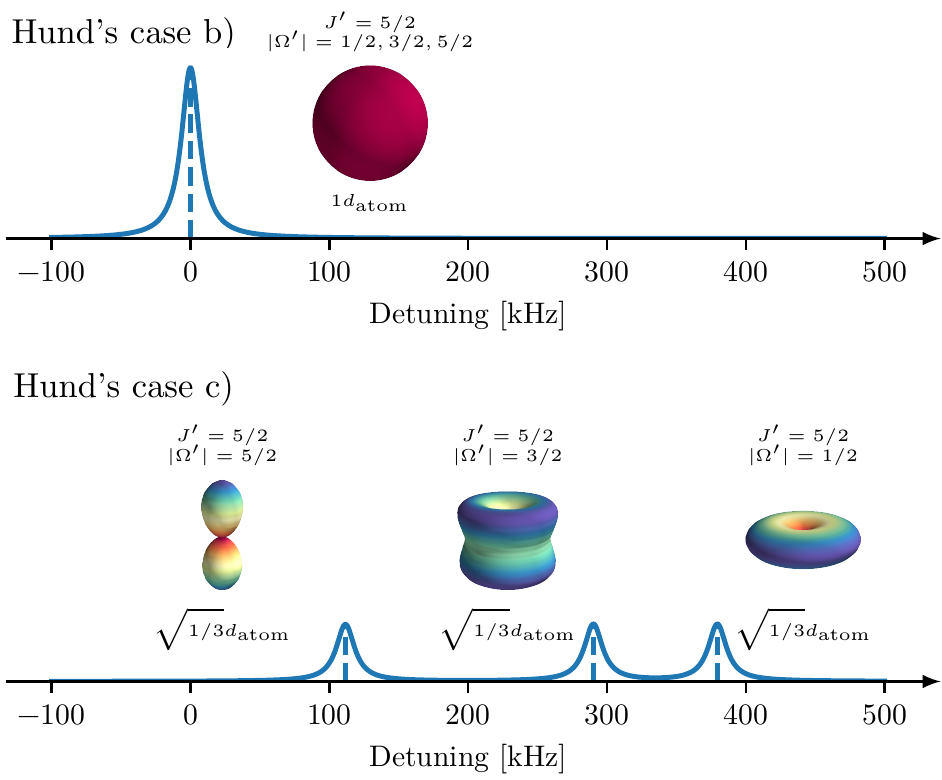}
	\caption{Molecular spectra and angular nuclear wave functions for the photoassociation of S-state Rydberg molecules in the $deep$ potential connected to the $(26\mathrm{S}_{1/2}+5\mathrm{S}_{1/2},F=2)$ manifold using a $\sigma^-$ transition from the intermediate state $\ket{6\mathrm{P}_{3/2},m_J=3/2,5\mathrm{S}_{1/2},F=2,m_F=+2}$. The spectra are the sum over individual Lorentz profiles with a width corresponding to the natural lifetime of the atomic Rydberg state. Each peak is scaled with the square of the transition matrix element. Zero detuning corresponds to the bare molecular state without rotational energy. Upper part: Hund's case b) coupling scheme, in which the rotational energy is zero. Lower part: Hund's case c) coupling scheme, in which the lines are shifted according to $B(J'(J'+1)-\Omega'^2)$ with $B=\SI{44}{\kHz}$. For reference, we give for each angular nuclear wave function the transition matrix element with respect to the atomic transition matrix element $d_\mathrm{atom}$ (excluding the Frank-Condon factor).}
\label{fig:26S12deep}		
\end{figure}

Let us first discuss the often used excitation from a fully stretched ground state $\ket{5\mathrm{S}_{1/2},F=2,m_F=+2,5\mathrm{S}_{1/2},F=2,m_F=+2}$ to a molecular S-state via two photons using a $\sigma^+$ transition for the lower and a $\sigma^-$ transition for the upper transition. In this case, the intermediate state is a pure fine structure state $\ket{6\mathrm{P}_{3/2},m_J=3/2,5\mathrm{S}_{1/2},F=2,m_F=+2}$ and corresponds in our framework to a total angular momentum of $J=7/2$ and a projection of the angular momentum onto the laboratory fixed z-axis of $m_J=7/2$. The only possible excitation for the second step is a $\sigma^-$ transition to states in the $deep$ potential connected to the $(26\mathrm{S}_{1/2} + 5\mathrm{S}_{1/2},F=2)$ manifold with $J'=5/2$ and $m_{J'}=5/2$.

As already mentioned earlier an ultra-long range S-state molecule is best described in a Hund's case b). In this description due to the vanishing spin-orbit interaction the spin states completely decouple from the molecular axis. The rotational energy is thus not given by eq.\,\eqref{equ:RotEnergy} but rather by
\begin{equation}
	E_\mathrm{rot} = B[P(P+1)-\Lambda^2]
\end{equation}
where $\vec P=\Lambda\vec{e}_{\overline{z}} + \vec N$ and $\Lambda$ is the projection of the orbital angular momentum of the electron $\vec L$ onto the molecular axis and equals zero for S-state molecules.

In the upper part of figure \ref{fig:26S12deep} this case is depicted. As the different $\Omega'$ states in a Hund's case c) description are degenerate, the excited molecular state is a superposition of all of them, resulting in a spherically symmetric nuclear wave function $\sum_{\Omega'} \left|D^{J'}_{m_J',\Omega'}(\alpha,\beta,0)\right|^2 = 1$. Note that this is only true as long as any residual spin-orbit coupling is weaker than the rotational energy. If this is not the case, the spins can be coupled to the internuclear axis. For S-state molecules, this can have two reasons: first, we know from the diagonalization of eq.\,\eqref{equ:hamiltonian} that even for molecular states with large bond length the scattering interaction mixes high-l states with non vanishing spin-orbit coupling into the potential curves. Second, relativistic effects in the triplet P-wave scattering \cite{Eiles2017SpinEffectsRydMol, markson2016SpinEffectsRydMol}, which are not included in this work, will also lead to a coupling to the molecular axis. If these couplings introduce a larger energy scale than the rotational constant, the molecule should be rather described in Hund's case c), as depicted in the lower part of figure \ref{fig:26S12deep}. Then, the resonance splits into three individual lines each with equal $|\Omega'|$ and the coupling leads to non-spherical angular nuclear wave functions $\left|D^{J'}_{m_J',|\Omega'|}(\alpha,\beta,0)\right|^2 + \left|D^{J'}_{m_J',-|\Omega'|}(\alpha,\beta,0)\right|^2$. We expect that in most cases an intermediate situation between Hund's case b) and c) is realized. However, irrespective of the coupling scheme within the molecule, the total transition rate summed over all rotational states is equal to the atomic one (excluding the Frank-Condon Factor). S-state Rydberg molecules can also be excited with a $\pi$-$\pi$ transition, however the transition matrix elements are then reduced by the additional Clebsch-Gordon coefficients for the two atomic transitions.


There is also an intuitive explanation within our framework why transitions to the corresponding $shallow$ potential are not accessible for the $\sigma^+ - \sigma^-$ excitation scheme. As the initial state is described by an S-wave ($N=0$) the intermediate state has the quantum numbers $J=K=7/2$. However from the numerical calculation of the excited potential, we know that it has a total electronic angular momentum of $K=3/2$. Thus, upon excitation, two quanta of electronic angular momentum have to be annihilated in the second absorption process. As the light field in dipole approximation can only carry away one quantum of angular momentum, the other one has to be transfered to the rotation of the molecule. Such a process is strongly suppressed or even absent due to the decoupling of the spin from the internuclear axis.

\begin{figure}
	\includegraphics[width=1\columnwidth]{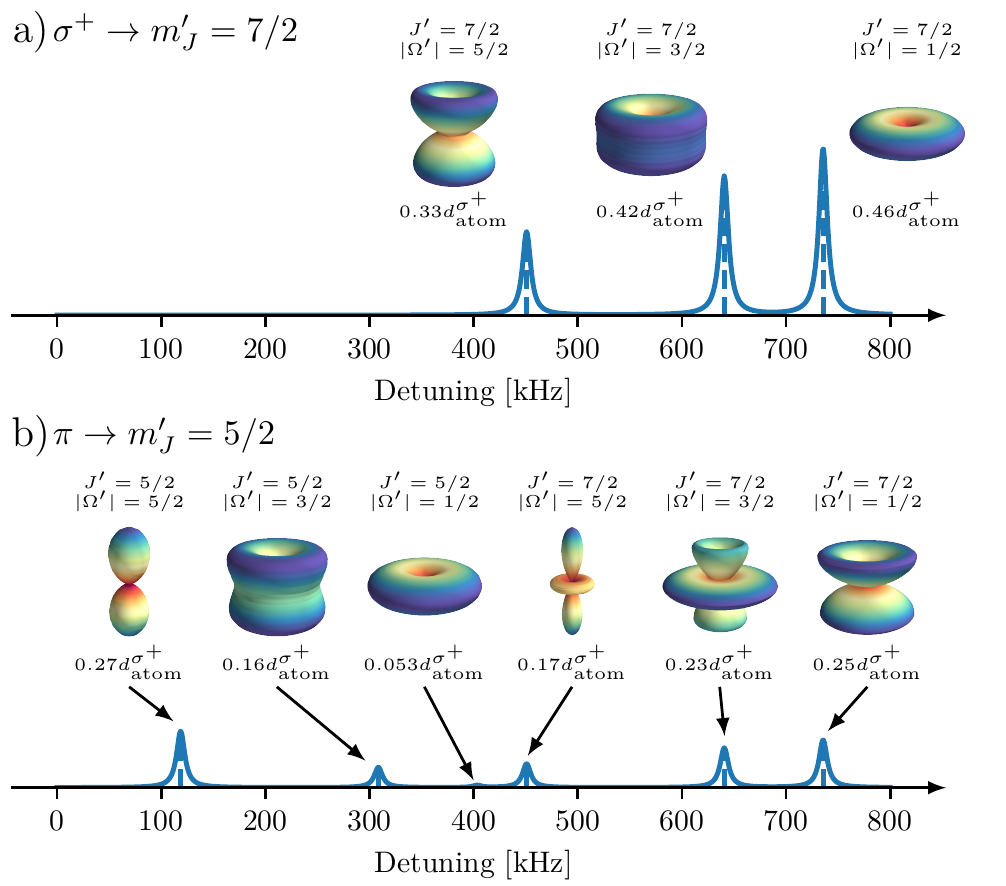}
	\caption{Molecular spectra for the photoassociation of P-state Rydberg molecules in the outermost well of the $deep$ potential connected to the $(25\mathrm{P}_{3/2}+5\mathrm{S}_{1/2},F=2)$ manifold from a fully stretched $\ket{5\mathrm{S}_{1/2},F=2,m_F=+2,5\mathrm{S}_{1/2},F=2,m_F=+2}$ initial state. The rotational constant is $B=\SI{47}{\kHz}$. a) for a $\sigma^+$ transition to $m_J'=7/2$ states b) for a $\pi$ transition to $m_J'=5/2$ states. For reference below the angular nuclear wave functions the matrix elements in reference to the atomic matrix element $d_\mathrm{atom}^{\sigma^+}$ for a $\sigma^+$ transition to the bare $\ket{25P_{3/2},m_J=3/2}$ state are stated. The two individual spectra are scaled in the same way.}
	\label{fig:25P32deep}
\end{figure}

Another important photoassociation scheme is the single photon transition to P-state Rydberg molecules. Because of the fine structure in the Rydberg state, a much richer molecular spectrum arises. Now, the spin-orbit interaction strongly couples the electronic spin to the internuclear axis. Exciting from a fully stretched ground state as before, $\ket{5\mathrm{S}_{1/2},F=2,m_F=+2,5\mathrm{S}_{1/2},F=2,m_F=+2}$, several rotational states in the $deep$ potential, connected to the $(25\mathrm{P}_{3/2}+5\mathrm{S}_{1/2}, F=2)$ manifold, can be excited. In figure \ref{fig:25P32deep} the molecular spectrum for a $\sigma^+$ and a $\pi$ transition are shown. As the initial state can be described through the quantum  numbers $J=5/2$, $m_J=5/2$, the excited state is either $m_J'=7/2$ or $m_J'=5/2$ respectively. In contrast to the previous example the coupling to different $\Omega'$ states is not of equal strength but rather depends on the individual state. Additionally for a $\pi$ transition $J'=5/2$ as well as $J'=7/2$ states have similar coupling strengths leading to rather complex excitation spectra. For narrow laser linewidth, this rotational structure should be observable in an experiment. 

\begin{figure}
	\includegraphics[width=1\columnwidth]{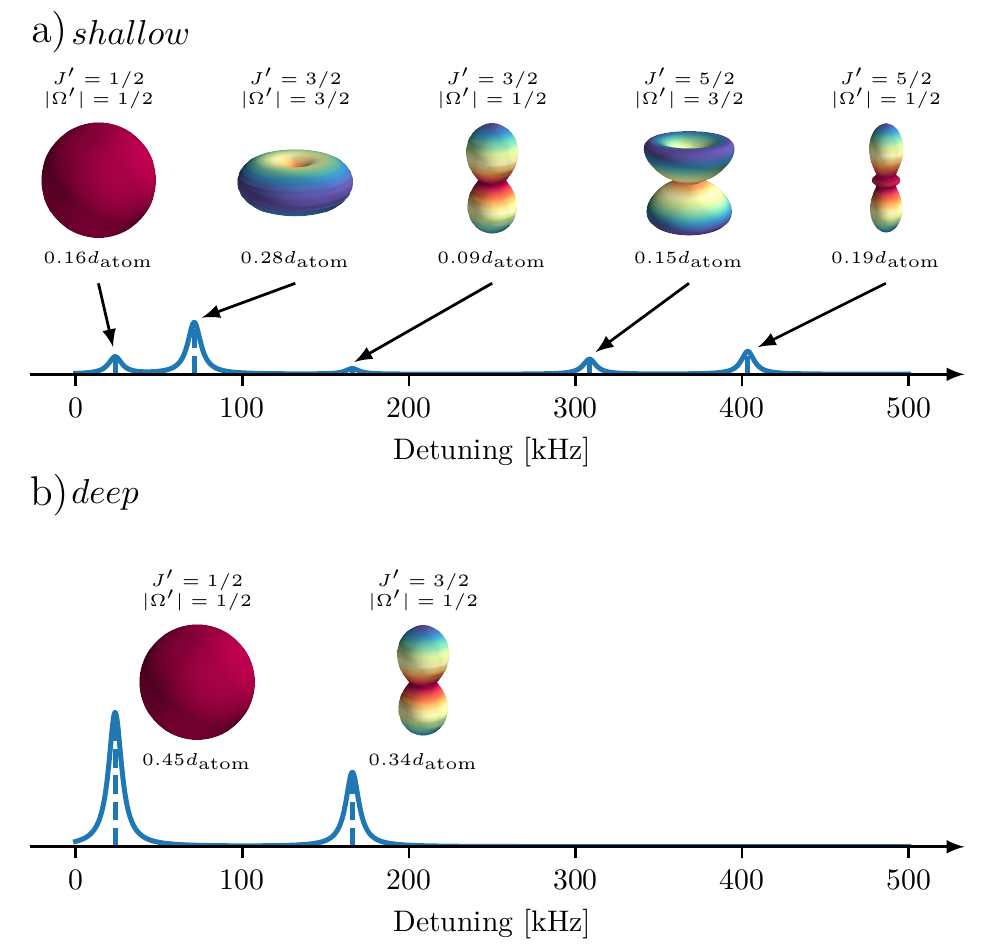}
	\caption{Molecular spectra and angular nuclear wave functions (as in figure \ref{fig:25P32deep}) for the excitation of the outermost well of the potentials connected to the $(25\mathrm{P}_{3/2}+5\mathrm{S}_{1/2}, F=1)$ manifold from a $\ket{5\mathrm{S}_{1/2},F=1,m_F=+1,5\mathrm{S}_{1/2},F=1,m_F=+1}$ initial state with a $\sigma^-$ transition. a) Excitation to the $shallow$ potential. b) Excitation to the $deep$ potential. Note that in both cases the excited states are not pure $m_J'$ states but rather superpositions of $m_J' = \pm 1/2$.}
	\label{fig:25P32F1}
\end{figure}

In figure \ref{fig:25P32F1} we show the molecular spectrum for a $\sigma^-$ transition from the $\ket{5\mathrm{S}_{1/2},F=1,m_F=+1,5\mathrm{S}_{1/2},F=1,m_F=+1}$ to the $deep$ and $shallow$ potentials of the $(25\mathrm{P}_{3/2}+5\mathrm{S}_{1/2},F=1)$ manifold. Again, a large variety of rotational states can be excited. Surprinsingly, we also find molecules with a spherical shape of the angular nuclear wave function, which one might not expect for a pure Hund's case c) coupling. This can be explained as follows. As already discussed before, different $|\Omega'|$ states now split due to the rotational energy. In addition, we know from the properties of the Wigner-D matrix that $\sum_{\Omega'=-J'}^{J'}\left|D^{J'}_{m_J',\Omega'}\right|^2=1$. Spherically symmetric molecules are therefore possible, if $\Omega'=\pm 1/2$ and $J'=1/2$. In the present example, this is the case for the $deep$ potential and the $shallow$ potential connected to an $F=1$ ground state manifold. In both cases, we have $K'=1/2$ and the lowest lying molecular state has a spherically symmetric nuclear wave function with $J'=1/2$ and $|\Omega'|=1/2$. This is rather unintuitive as the underlying electronic wave function is off course non spherically symmetric due to the coupling of the electronic orbital wave function to the internuclear axis.



\section{Conclusion}

To conclude, we have presented a complete treatment of ultra-long range molecular Rydberg states including rotation in a laboratory fixed frame. We have calculated transition matrix elements for the photoassociation of Rydberg molecules and discussed the implications for the excited spectra and angular nuclear wave functions for selected transitions. Our results can be easily transfered to other Rydberg systems like Trilobite or Butterfly molecules and atomic species.

The formalism described in this work is not limited to this kind of application. In a similar way transitions between different molecular states can be calculated and even interactions between molecular states can be investigated. As the description is rather generic it should be straighforward to adapt it for cases including spin dependent scattering interactions not discussed here \cite{Eiles2017SpinEffectsRydMol, markson2016SpinEffectsRydMol} or generalizing it further, fully including the hyperfine interaction of the Rydberg state. With state of the art experiments the presented features should be measurable in typical spectroscopic experiments in ultracold systems.

\section{Acknowledgements}

We would like to thank M. Lemeshko for helpful discussions and I. Fabrikant for providing the $e^-$-Rb scattering phase-shifts used to calculate the molecular potentials. H.O. acknowledges financial support by the DFG within the SFB/TR 49 and the SFB/TR 185. C.L. acknowledges financial support by the DFG within the SFB/TR 185. O.T. acknowledges financial support by the DFG within the SFB/TR 49 and the MAINZ graduate school.

\end{document}